\documentstyle[aps]{revtex}

\begin{document}
\draft
\title{Casimir effect for the massless Dirac field in two-dimensional
Reissner-Nordstr\"{o}m spacetime}
\author{Xin-Qin Gao\thanks{%
Email: xinqin\_gao@yahoo.com.cn} and Jian-Yang Zhu\thanks{%
Email: zhujy@bnu.edu.cn}}
\address{Department of Physics, Beijing Normal University, Beijing, 100875,China}
\maketitle

\begin{abstract}
In this paper, the two-dimensional Reissner-Nordstr\"{o}m black hole is
considered as a system of the Casimir type. In this background the Casimir
effect for the massless Dirac field is discussed. The massless Dirac field
is confined between two ``parallel plates'' separated by a distance $L$ and
there is no particle current drilling through the boundaries. The vacuum
expectation values of the stress tensor of the massless Dirac field at
infinity are calculated separately in the Boulware state, the Hartle-Hawking
state and the Unruh state.

\end{abstract}


\section{Introduction}

In 1948, Casimir first predicted that two neutral parallel plates in vacuum
would attract each other \cite{Casimir}, a phenomenon later known as the
Casimir effect. This effect is a manifestation of the non-trivial properties
of the vacuum state in quantum field theory, and it is a macroscopic quantum
effect.

The Casimir effect is an interdisciplinary subject. It has been studied in
many fields of physics, such as quantum field theory, atomic and molecular
physics, condensed matter physics, gravitation and cosmology \cite{Bordag}.

Since the first prediction, people have investigated cases of various
geometries, boundaries, fields and backgrounds \cite{Kimbal}. With improved
experiment techniques, a lot of theoretical predictions have been verified,
arousing even more interest.

When one studies the Casimir effect, the zero point energy of the confined
system should be calculated, and regularized \cite{Plunie,Milton}. There are
many methods of regularization \cite{Deutsc,Hawking,Capper,Christ}\ such as
Green's function method, zeta function regularization, dimensional
regularization, and point-splitting method, etc. The zero point energy after
regularization is divergent, and thus it needs renormalization, which,
briefly, aims at shifting the divergent part of the ground state energy from
the ground state energy to the classical energy \cite{Bordag}, and can be
written as 
\[
E=E^{class}+E_0=\left( E^{class}+E_0^{div}\right) +\left(
E_0-E_0^{div}\right) =\tilde{E}^{class}+E_0^{ren}. 
\]

In order to study the Casimir effect in a curved spacetime, the vacuum
expectation value of the energy-momentum tensor needs to be calculated. But
it is very difficult to calculate the energy-momentum tensor directly using
the formula 
\[
T_{\mu \nu }=\frac 2{\sqrt{-g}}\frac{\delta S_m}{\delta g^{\mu \nu }}, 
\]
especially in higher dimensions. In Refs \cite{Setare,Chris,Vage}, Wald's
axioms \cite{Wald-1,Wald-2,Davies} for renormalized energy-momentum tensor
were used to calculate the Casimir effect indirectly in two-dimensional
curved spacetime. The massless scalar field in two-dimensional Schwarzschild
black hole background \cite{Setare}, two-dimensional string black hole
background \cite{Chris}, and two-dimensional Achucarro-Ortiz black hole
background \cite{Vage} were studied, respectively. On the one hand, as far
as we know, the Casimir effect for the Dirac field in curved spacetime is
yet to be studied. On the other hand, we know that the energy-momentum
tensor in four-dimensional black hole background could be calculated in some
cases, for example in Refs. \cite{Shin,Odin}, but there is no boundary to
confine the field in those works. The energy-momentum tensor for a field
confined by some boundary is very difficult to calculate in curved
spacetime. In this paper we will study the Casimir effect for the massless
Dirac field using Wald's axioms, however the background spacetime is chosen
to be two-dimensional Reissner-Nordstr\"{o}m (RN) black hole, not
four-dimensional background. We want to know if there are some differences
in Casimir energy momentum tensor and Casimir force between scalar field and
massless Dirac field in the same background. We also research the depedent
relationship of Casimir force to background and massless field.

Wald's axioms are as follows: (1) Expectation values of the energy-momentum
tensor are covariantly conserved. (2) Causality holds. (3) Standard results
in Minkowski spacetime can be obtained. (4) Standard results for the
`off-diagonal' elements can be obtained. (5) The energy-momentum tensor is a
local functional of the metric; i.e., it depends only on the metric and its
derivatives which appear through the Riemann curvature tensor and the
metric's covariant derivatives up to the second order. Here condition (3)
means that the normal ordering procedure in the Minkowski spacetime should
be valid, and condition (4) is simply the observation that, as $\left\langle
\Phi \left| T_{\mu \nu }\right| \Psi \right\rangle $ is in any case finite
for orthogonal states, $\left\langle \Phi \mid \Psi \right\rangle =0$, this
quantity should be of the usual value\cite{Davies}.

Due to the existence of the horizon and the effective potential barrier, the
two-dimensional RN black hole is considered as a system of the Casimir type.
Applying Wald's axioms\ for the energy-momentum tensor, and the anomaly
trace of the energy-momentum tensor, the regularized energy-momentum tensor
of the massless Dirac field in two-dimensional RN background is calculated
and the Casimir effect is discussed in this paper. The organization is as
follows. In Sec. \ref{Sec. 2}, the background spacetime is given and the
most general form of the energy-momentum tensor in the given background is
obtained. Then in Sec. \ref{Sec. 3}, the renormalized energy-momentum tensor
for the massless Dirac field in the two-dimensional Minkowski spacetime is
calculated. In Sec. \ref{Sec. 4}, the expectation values of the renormalized
energy-momentum tensor for the massless Dirac field in the Boulware vacuum,
the Hartle-Hawking vacuum and the Unruh vacuum are calculated respectively,
and from that the Casimir force acting on two ``parallel plates''\ in the
given background is extracted. Finally, Sec. \ref{Sec. 5} contains the
summarization and the conclusions.

\section{Background spacetime and the general description of the
energy-momentum tensor}

\label{Sec. 2}

The RN spacetime is the background of this paper. The line element of a
two-dimensional RN black hole is 
\begin{equation}
ds^2=\left( 1-\frac{2m}r+\frac{Q^2}{r^2}\right) dt^2-\left( 1-\frac{2m}r+%
\frac{Q^2}{r^2}\right) ^{-1}dr^2,  \label{line}
\end{equation}
where $m$ and $Q$ are the mass and the charge of the black hole respectively.

In order to make the best use of the conclusions in the Minkowski spacetime,
metric (\ref{line}) is written in the following form under conformal
transformation 
\begin{equation}
ds^2=f(r)\left( dt^2-du^2\right) ,  \label{conformal}
\end{equation}
in which 
\begin{equation}
f(r)=1-\frac{2m}r+\frac{Q^2}{r^2},\frac{dr}{du}=f(r).  \label{f(r)}
\end{equation}
The non-zero Christoffel symbols of metric (\ref{conformal}) are 
\begin{equation}
\Gamma _{tt}^u=\Gamma _{tu}^t=\Gamma _{uu}^t=\Gamma _{uu}^u=\frac 12\frac{%
df(r)}{dr}=\frac{mr-Q^2}{r^3}.  \label{Chr}
\end{equation}
The scalar curvature of metric (\ref{conformal}) is 
\begin{equation}
R=\frac{-2\left( 2mr-3Q^2\right) }{r^4}.  \label{scalar}
\end{equation}
The conservation equation of energy-momentum tensor $\nabla _a\langle
T_b^a\rangle _{ren}=0$ can be extended as \cite{Wald-3} 
\begin{equation}
\partial _u\langle T_t^u\rangle _{ren}+\Gamma _{tu}^t\langle T_t^u\rangle
_{ren}-\Gamma _{tt}^u\langle T_u^t\rangle _{ren}=0,  \label{con1}
\end{equation}
\begin{equation}
\partial _u\langle T_u^u\rangle _{ren}+\Gamma _{tu}^t\langle T_u^u\rangle
_{ren}-\Gamma _{tu}^t\langle T_t^t\rangle _{ren}=0,  \label{con2}
\end{equation}
where 
\begin{equation}
\langle T_u^t\rangle _{ren}=-\langle T_t^u\rangle _{ren},\left\langle
T_t^t\right\rangle _{ren}=\left\langle T_\alpha ^\alpha \right\rangle
_{ren}-\langle T_u^u\rangle _{ren},  \label{antisym}
\end{equation}
in which $\left\langle T_\alpha ^\alpha \right\rangle _{ren}$ is the anomaly
trace \cite{Chr}.

References \cite{Davie,Deser} tell us that the anomaly trace of the
energy-momentum tensor in two dimensions of a massless Dirac field after
renormalization is the same as a scalar field 
\begin{equation}
\langle T_\alpha ^\alpha \rangle _{ren}=-\frac R{24\pi }.  \label{traceall}
\end{equation}
In the RN background, applying Eq. (\ref{scalar}), Eq. (\ref{traceall})
becomes 
\begin{equation}
\langle T_\alpha ^\alpha \rangle _{ren}=-\frac 1{24\pi }\frac{2\left(
2mr-3Q^2\right) }{r^4}.  \label{trace}
\end{equation}
Equations (\ref{Chr}) and (\ref{antisym}) being used, Eq. (\ref{con1})
becomes 
\begin{equation}
\frac \partial {\partial r}\left( f(r)\left\langle T_t^u\right\rangle
_{ren}\right) =0,  \label{con11}
\end{equation}
The solution of Eq. (\ref{con11}) is 
\begin{equation}
\langle T_t^u\rangle _{ren}=\alpha f^{-1}(r),  \label{con111}
\end{equation}
where $\alpha $ is a constant of integration. In the same way Eq. (\ref{con2}%
) becomes 
\begin{equation}
\frac \partial {\partial r}\left( f(r)\langle T_u^u\rangle _{ren}\right) =%
\frac 12\left( \frac{df(r)}{dr}\right) \left\langle T_\alpha ^\alpha
\right\rangle _{ren}.  \label{con22}
\end{equation}
The solution of Eq. (\ref{con22}) is 
\begin{equation}
\langle T_u^u(r)\rangle _{ren}=\left[ H(r)+\beta\right] f^{-1}(r),
\label{con222}
\end{equation}

in which $\beta $ is a constant of integration and 
\begin{equation}
H(r)=\frac 12\int_{r_H}^r\left\langle T_u^u(r^{\prime })\right\rangle _{ren}%
\frac d{dr^{\prime }}f(r^{\prime })dr^{\prime }.  \label{Hr}
\end{equation}
Substituting Eqs. (\ref{f(r)}) and (\ref{trace}) to Eq. (\ref{Hr}), we get 
\begin{eqnarray}
H(r) &=&-\frac 1{24\pi }\left( \frac{m^2}{r^4}-\frac{2mQ^2}{r^5}+\frac{Q^4}{%
r^6}\right)  \nonumber \\
&&+\frac 1{24\pi }\left[ \frac{m^2}{\left( m+\sqrt{m^2-Q^2}\right) ^4}-\frac{%
2mQ^2}{\left( m+\sqrt{m^2-Q^2}\right) ^5}+\frac{Q^4}{\left( m+\sqrt{m^2-Q^2}%
\right) ^6}\right] .  \label{H}
\end{eqnarray}
Using Eqs. (\ref{antisym}), (\ref{con111}) and (\ref{con222}), we get the
general expression of the energy-momentum tensor for a massless Dirac field
in the given two-dimensional RN background 
\begin{equation}
\left\langle T_\nu ^\mu \right\rangle _{ren}=\left( 
\begin{array}{ll}
\left\langle T_\alpha ^\alpha \right\rangle _{ren}-f^{-1}\left( r\right) H(r)
& 0 \\ 
0 & f^{-1}\left( r\right) H(r)
\end{array}
\right) +f^{-1}(r)\left( 
\begin{array}{ll}
-\beta & -\alpha \\ 
\alpha & \beta
\end{array}
\right) .  \label{T}
\end{equation}

In expression (\ref{T}) the values of the undetermined constants $\alpha $
and $\beta $ depend on which vacuum state the massless Dirac field is in.
The energy-momentum tensor can have different expression in different vacuum
states. In the following we will determine the value of $\alpha $ and $\beta 
$ by imposing Wald's axioms (3) and (4). In this way, we will work out the
energy-momentum tensors for a massless Dirac field in the given
two-dimensional RN background in three vacuum states.

\section{Renormalized energy-momentum tensor for a massless Dirac field in
the Minkowski spacetime}

\label{Sec. 3}

Two ``parallel plates''\ are placed at $r_1$ and $r_2=r_1+L$. The massless
Dirac field is confined in the interval of $r_1<r<r_1+L.$ The physical
boundary condition is that there is no particle current through the
``walls'', i.e., $\hat{n}\cdot \vec{j}\left( x\right) =0$ at $r_1$ and $%
r_2=r_1+L$, where $\hat{n}$ is the unit vector normal to the boundary
surface. $\hat{n}$ equals $\hat{r}$ at $r_1$, and $\hat{n}$ equals $-\hat{r}$
at $r_2=r_1+L$.

To satisfy this boundary condition, the momentum of a massless Dirac field
in the confined direction should be \cite{Milonn} 
\begin{equation}
p=p\left( n,L\right) =\left( n+\frac 12\right) \frac \pi L.
\end{equation}
Thus the zero point energy of the massless Dirac field satisfying the given
boundary condition is 
\begin{equation}
E_B=-\sum_{n=0}^\infty \left( n+\frac 12\right) \frac \pi L.
\end{equation}
We can easily write down the zero point energy of the massless Dirac field
without boundary 
\begin{equation}
E_0=-\frac 1{2\pi }\int_0^\infty kdk,
\end{equation}
in which $k$ is the wave vector.

Thus, the Casimir energy for the massless Dirac field is 
\begin{eqnarray}
E_c &=&E_B-E_0  \nonumber \\
&=&-\sum_{n=0}^\infty \left( n+\frac 12\right) \frac \pi L+\frac 1{2\pi }%
\int_0^\infty kdk  \nonumber \\
&=&-\frac \pi L\left[ \sum_{n=0}^\infty \left( n+\frac 12\right) -\left( 
\frac L{\sqrt{2}\pi }\right) ^2\int_0^\infty kdk\right]  \nonumber \\
&=&-\frac \pi L\left[ \sum_{n=0}^\infty \left( n+\frac 12\right)
-\int_0^\infty tdt\right] ,  \label{E1}
\end{eqnarray}
in which $t=\frac L{\sqrt{2}\pi } k$. We define $F\left( t\right) =t$, and
Eq. (\ref{E1}) becomes 
\begin{equation}
E_c=-\frac \pi L\left[ \sum_{n=0}^\infty F\left( n+\frac 12\right)
-\int_0^\infty F\left( t\right) dt\right] .
\end{equation}
Applying the Abel-Plana formula \cite{Mostep} 
\begin{equation}
\sum_{n=0}^\infty F\left( n+\frac 12\right) -\int_0^\infty dtF\left(
t\right) =-i\int_0^\infty \frac{dt}{e^{2\pi t}+1}\left[ F\left( it\right)
-F\left( -it\right) \right] ,
\end{equation}
we get 
\begin{eqnarray}
E_c &=&-\frac \pi L\left( -i\right) \int_0^\infty \frac{dt}{e^{2\pi t}+1}%
\left[ F\left( it\right) -F\left( -it\right) \right]  \nonumber \\
&=&-\frac{2\pi }L\int_0^\infty \frac{tdt}{e^{2\pi t}+1}  \nonumber \\
&=&\frac{-\pi }{24L}.  \label{Ec}
\end{eqnarray}
So the Casimir energy density is 
\begin{equation}
\rho =\frac{E_c}L=\frac{-\pi }{24L^2}.  \label{p}
\end{equation}
Equation (\ref{p}) is the same as the result of Ref. \cite{Paola} in which $%
D=2$. Therefore we obtain the standard Casimir energy-momentum tensor of the
massless Dirac field confined by two ``parallel plates'' in the $1+1$
dimensional Minkowski spacetime 
\begin{equation}
\left\langle T_\nu ^\mu \right\rangle _{ren}=\left( 
\begin{array}{cc}
\frac{-\pi }{24L^2} & 0 \\ 
0 & \frac \pi {24L^2}
\end{array}
\right) =\frac \pi {24L^2}\left( 
\begin{array}{cc}
-1 & 0 \\ 
0 & 1
\end{array}
\right) ,  \label{TC}
\end{equation}
which is identical to the result of the massless scalar field in the
two-dimensional flat spacetime with the same constraint condition \cite{Kimb}%
.

\section{Renormalized energy-momentum tensor for a massless Dirac field in
three vacuum states}

\label{Sec. 4}

We consider the two-dimensional RN black hole as a system of the Casimir
type. Because the two-dimensional RN black hole is asymptotically flat at
infinity, we require the renormalized energy-momentum tensor (\ref{T}) in
the two-dimensional RN background at infinity be equal to the standard
Casimir energy-momentum tensor in the Minkowski spacetime (\ref{TC}). In
this way $\alpha $ and $\beta $ are determined. Then we get the Casimir
energy-momentum tensors for the massless Dirac field under the given
boundary condition in the two-dimensional RN background in three vacuum
states separately.

\subsection{Boulware vacuum}

The Boulware vacuum (denoted by $\left| B\right\rangle $) \cite{Boulwa}
assumes that there is no particle at infinity (towards ${\cal J}^{+}$).
Comparing Eq. (\ref{TC}) with Eq. (\ref{T}) at $r\rightarrow \infty $, we
get 
\begin{equation}
\alpha =0,  \label{gg1}
\end{equation}
and 
\begin{equation}
\beta =\frac \pi {24L^2}-\frac 1{24\pi }\left[ \frac{m^2}{\left( m+\sqrt{%
m^2-Q^2}\right) ^4}-\frac{2mQ^2}{\left( m+\sqrt{m^2-Q^2}\right) ^5}+\frac{Q^4%
}{\left( m+\sqrt{m^2-Q^2}\right) ^6}\right] .  \label{gg2}
\end{equation}

Substituting expressions (\ref{gg1}) and (\ref{gg2}) into the general
expression of the energy-momentum tensor (\ref{T}), we get 
\begin{eqnarray}
\left\langle B\left| T_\nu ^\mu \right| B\right\rangle _{ren} &=&\left( 
\begin{array}{cc}
\left\langle T_\alpha ^\alpha \right\rangle _{ren}-f^{-1}(r)H(r) & 0 \\ 
0 & f^{-1}(r)H(r)
\end{array}
\right) +f^{-1}\left( r\right) \left\{ \frac \pi {24L^2}-\frac 1{24\pi }%
\right.  \nonumber \\
&&\left. \times \left[ \frac{m^2}{\left( m+\sqrt{m^2-Q^2}\right) ^4}-\frac{%
2mQ^2}{\left( m+\sqrt{m^2-Q^2}\right) ^5}+\frac{Q^4}{\left( m+\sqrt{m^2-Q^2}%
\right) ^6}\right] \right\}  \nonumber \\
&&\times \left( 
\begin{array}{cc}
-1 & 0 \\ 
0 & 1
\end{array}
\right) .  \label{TB}
\end{eqnarray}
This expression is the renormalized energy-momentum tensor for a confined
massless Dirac field in the given two-dimensional RN background in the
Boulware vacuum.

\subsection{Hartle-Hawking vacuum}

In the Hartle-Hawking vacuum (denoted by $\left| H\right\rangle $) \cite
{Hartle}, the black hole is considered to be in a thermal equilibrium state
at Hawking temperature $T$, and there is an infinite heat reservoir of black
body radiation around the black hole. In order to calculate the
energy-momentum tensor of the gas in a thermal equilibrium state, the system
is considered as a canonical system. The number of microscopic states is 
\begin{eqnarray}
D\left( \varepsilon \right) d\varepsilon &=&g_s\frac 1h\frac{d\sum \left(
\varepsilon \right) }{d\varepsilon }d\varepsilon =g_s\frac 1h\frac d{%
d\varepsilon }\left( \int dxdp\right) d\varepsilon  \nonumber \\
&=&g_s\frac Lh\left( \frac d{d\varepsilon }\int d\varepsilon \right)
d\varepsilon =g_s\frac L{hc}d\varepsilon .  \label{number}
\end{eqnarray}
The particle population in the energy interval $\varepsilon \rightarrow
\varepsilon +d\varepsilon $ is 
\begin{equation}
dn=n_\varepsilon D\left( \varepsilon \right) d\varepsilon =\frac{D\left(
\varepsilon \right) d\varepsilon }{e^{\beta \varepsilon }+1}.  \label{dn}
\end{equation}
From Eqs. (\ref{number}) and (\ref{dn}), the energy density of black body
radiation in one dimensional space is 
\begin{equation}
\rho \left( \varepsilon ,T\right) d\varepsilon =\frac{\varepsilon dn}L=g_s%
\frac 1{hc}\frac{\varepsilon d\varepsilon }{e^{\beta \varepsilon }+1}=g_s%
\frac 1{hc}\frac{\hbar \omega \hbar d\omega }{e^{\frac{\hbar \omega }{K_BT}%
}+1}.  \label{de}
\end{equation}
In natural units, Eq. (\ref{de}) becomes 
\begin{equation}
\rho \left( \varepsilon ,T\right) d\varepsilon =\frac{g_s}{2\pi }\frac{%
\omega d\omega }{e^{\frac \omega T}+1}.  \label{den}
\end{equation}
Thus in two-dimensional flat spacetime, the energy density of a massless
Dirac field in a thermal equilibrium state is 
\begin{equation}
\left\langle T_{tt}\right\rangle =\int \frac{g_s}{2\pi }\frac{\omega d\omega 
}{e^{\frac \omega T}+1}=g_s\frac \pi {24}T^2=\frac \pi {24}T^2\times 2.
\label{Ttt}
\end{equation}
There is no momentum flow in a thermal equilibrium state at temperature $T$,
and thus the energy-momentum tensor of the gas in a thermal equilibrium
state can be written as 
\begin{equation}
\left\langle H\left| T_\nu ^\mu \right| H\right\rangle =\frac \pi {24}%
T^2\left( 
\begin{array}{cc}
2 & 0 \\ 
0 & -2
\end{array}
\right) =-\frac \pi {12}T^2\left( 
\begin{array}{cc}
-1 & 0 \\ 
0 & 1
\end{array}
\right) .  \label{TH}
\end{equation}

In the Hartle-Hawking vacuum, the standard Casimir energy-momentum tensor (%
\ref{TC}) is modified by Eq. (\ref{TH}) to 
\begin{equation}
\left\langle H\left| T_\nu ^\mu \right| H\right\rangle _{ren}=\frac \pi {%
24L^2}\left( 
\begin{array}{cc}
-1 & 0 \\ 
0 & 1
\end{array}
\right) -\frac \pi {12}T^2\left( 
\begin{array}{cc}
-1 & 0 \\ 
0 & 1
\end{array}
\right) ,  \label{T+H}
\end{equation}
in which $T$ is the Hawking temperature of the two-dimensional RN black
hole. The temperature $T$ is the same as the temperature in the four
dimensional RN black hole \cite{Zhu}. Therefore 
\begin{equation}
T=\frac 1{2\pi }\frac{\sqrt{m^2-Q^2}}{\left( m+\sqrt{m^2-Q^2}\right) ^2}.
\label{T11111}
\end{equation}
Substituting Eq. (\ref{T11111}) into Eq. (\ref{T+H}), we get the
renormalized energy-momentum tensor for a massless Dirac field in the two
dimensional RN background in the Hartle-Hawking vacuum at infinity (towards $%
{\cal J}^{+}$) 
\begin{equation}
\left\langle H\left| T_\nu ^\mu \right| H\right\rangle _{ren}=\frac \pi {%
24L^2}\left( 
\begin{array}{cc}
-1 & 0 \\ 
0 & 1
\end{array}
\right) -\frac \pi {12}\frac 1{4\pi ^2}\frac{m^2-Q^2}{\left( m+\sqrt{m^2-Q^2}%
\right) ^4}\left( 
\begin{array}{cc}
-1 & 0 \\ 
0 & 1
\end{array}
\right) .  \label{2}
\end{equation}
Expression (\ref{2}) corresponds to (\ref{T}) at $r\rightarrow \infty $, we
get the values of $\alpha $ and $\beta $ 
\begin{equation}
\alpha =0,  \label{a111}
\end{equation}
\begin{equation}
\beta =\frac \pi {24L^2}-\frac 1{24\pi }\left[ \frac{\frac 32m^2-\frac 12Q^2%
}{\left( m+\sqrt{m^2-Q^2}\right) ^4}-\frac{2mQ^2}{\left( m+\sqrt{m^2-Q^2}%
\right) ^5}+\frac{Q^4}{\left( m+\sqrt{m^2-Q^2}\right) ^6}\right] .  \label{b}
\end{equation}

Substituting expressions (\ref{a111}) and (\ref{b}) into the general
expression of the energy-momentum tensor (\ref{T}), we get 
\begin{eqnarray}
\left\langle H\left| T_\nu ^\mu \right| H\right\rangle _{ren} &=&\left( 
\begin{array}{cc}
\left\langle T_\alpha ^\alpha \right\rangle _{ren}-f^{-1}(r)H(r) & 0 \\ 
0 & f^{-1}(r)H(r)
\end{array}
\right) +f^{-1}(r)\times \left\{ \frac \pi {24L^2}-\frac 1{24\pi }\right. 
\nonumber \\
&&\left. \times \left[ \frac{\frac 32m^2-\frac 12Q^2}{\left( m+\sqrt{m^2-Q^2}%
\right) ^4}-\frac{2mQ^2}{\left( m+\sqrt{m^2-Q^2}\right) ^5}+\frac{Q^4}{%
\left( m+\sqrt{m^2-Q^2}\right) ^6}\right] \right\}  \nonumber \\
&&\times \left( 
\begin{array}{cc}
-1 & 0 \\ 
0 & 1
\end{array}
\right) .  \label{v}
\end{eqnarray}
This expression is the renormalized energy-momentum tensor for a confined
massless Dirac field in the given two-dimensional RN background in the
Hartle-Hawking vacuum state.

\subsection{Unruh vacuum}

The Unruh vacuum (denoted by $\left| U\right\rangle $) is considered as a
vacuum in which the two-dimensional RN black hole is in an equilibrium state
at Hawking temperature $T$ \cite{Unruh}, but due to the Hawking radiation,
massless particles are detected at infinity (towards ${\cal J}^{+}$) in this
vacuum state. The energy-momentum tensor has not only energy density
component but also momentum density component. The energy density of the
massless black body radiation spreading in the $u$ direction is half of that
in Eq. (\ref{Ttt}). Thus the energy-momentum tensor in the limit $%
r\rightarrow \infty $ and in the absence of the boundary conditions is as
follows 
\begin{equation}
\left\langle U\left| T_\nu ^\mu \right| U\right\rangle =\frac \pi {24}%
T^2\left( 
\begin{array}{cc}
1 & 1 \\ 
-1 & -1
\end{array}
\right) .  \label{U}
\end{equation}
This means that the standard Casimir energy-momentum tensor of a massless
Dirac field (\ref{TC}) must be modified by Eq. (\ref{U}) to 
\begin{equation}
\left\langle U\left| T_\nu ^\mu \right| U\right\rangle _{ren}=\frac \pi {%
24L^2}\left( 
\begin{array}{cc}
-1 & 0 \\ 
0 & 1
\end{array}
\right) -\frac \pi {24}T^2\left( 
\begin{array}{cc}
-1 & -1 \\ 
1 & 1
\end{array}
\right) ,  \label{TU}
\end{equation}
where $T$ is the Hawking temperature of the two-dimensional RN black hole.
Substituting expression (\ref{T11111}) into (\ref{TU}) we get the
renormalized energy-momentum tensor for a massless Dirac field in the two
dimensional RN background in the Unruh vacuum at infinity (towards ${\cal J}%
^{+}$) 
\begin{equation}
\left\langle U\left| T_\nu ^\mu \right| U\right\rangle _{ren}=\frac \pi {%
24L^2}\left( 
\begin{array}{cc}
-1 & 0 \\ 
0 & 1
\end{array}
\right) -\frac \pi {24}\frac 1{4\pi ^2}\frac{m^2-Q^2}{\left( m+\sqrt{m^2-Q^2}%
\right) ^4}\left( 
\begin{array}{cc}
-1 & -1 \\ 
1 & 1
\end{array}
\right) .  \label{h}
\end{equation}
Expression (\ref{h}) corresponds to (\ref{T}) at $r\rightarrow \infty $, and
in this way we determine the values of $\alpha $ and $\beta $ 
\begin{equation}
\alpha =\frac 1{96\pi }\frac{m^2-Q^2}{\left( m+\sqrt{m^2-Q^2}\right) ^4},
\label{aa}
\end{equation}
\begin{equation}
\beta =\frac \pi {24L^2}-\frac 1{24\pi }\left[ \frac{\frac 54m^2-\frac 14Q^2%
}{\left( m+\sqrt{m^2-Q^2}\right) ^4}-\frac{2mQ^2}{\left( m+\sqrt{m^2-Q^2}%
\right) ^5}+\frac{Q^4}{\left( m+\sqrt{m^2-Q^2}\right) ^6}\right] .
\label{bb}
\end{equation}

Substituting expressions (\ref{aa}) and (\ref{bb}) into the general
expression of the energy-momentum tensor (\ref{T}), we get 
\begin{eqnarray}
\left\langle U\left| T_\nu ^\mu \right| U\right\rangle _{ren} &=&\left( 
\begin{array}{cc}
\left\langle T_\alpha ^\alpha \right\rangle _{ren}-f^{-1}(r)H(r) & 0 \\ 
0 & f^{-1}(r)H(r)
\end{array}
\right) +f^{-1}(r)\times  \nonumber \\
&&\left( 
\begin{array}{c}
-\frac \pi {24L^2}+\frac 1{24\pi }\left[ \frac{\frac 54m^2-\frac 14Q^2}{%
\left( m+\sqrt{m^2-Q^2}\right) ^4}-\frac{2mQ^2}{\left( m+\sqrt{m^2-Q^2}%
\right) ^5}+\frac{Q^4}{\left( m+\sqrt{m^2-Q^2}\right) ^6}\right] \\ 
\frac 1{96\pi }\frac{m^2-Q^2}{\left( m+\sqrt{m^2-Q^2}\right) ^4}
\end{array}
\right.  \nonumber \\
&&\left. 
\begin{array}{c}
-\frac 1{96\pi }\frac{m^2-Q^2}{\left( m+\sqrt{m^2-Q^2}\right) ^4} \\ 
\frac \pi {24L^2}-\frac 1{24\pi }\left[ \frac{\frac 54m^2-\frac 14Q^2}{%
\left( m+\sqrt{m^2-Q^2}\right) ^4}-\frac{2mQ^2}{\left( m+\sqrt{m^2-Q^2}%
\right) ^5}+\frac{Q^4}{\left( m+\sqrt{m^2-Q^2}\right) ^6}\right]
\end{array}
\right) .  \label{Z}
\end{eqnarray}
This expression is the renormalized energy-momentum tensor for a confined
massless Dirac field in the given two-dimensional RN background in the Unruh
vacuum state.

\subsection{Casimir force}

Expressions (\ref{TB}), (\ref{v}) and (\ref{Z}) are the renormalized
energy-momentum tensors for a confined massless Dirac field in the given
two-dimensional RN background in the Boulware\ vacuum, the Hartle-Hawking
vacuum and the Unruh vacuum respectively. In expressions (\ref{TB}), (\ref{v}%
) and (\ref{Z}), 
\begin{equation}
\frac \pi {24L^2}f^{-1}(r)\left( 
\begin{array}{cc}
-1 & 0 \\ 
0 & 1
\end{array}
\right)  \label{boundary}
\end{equation}
is the contribution to vacuum fluctuation due to the presence of boundaries.
The energy-momentum tensors in three vacuum states (\ref{TB}), (\ref{v}) and
(\ref{Z}) are separable as follows: 
\begin{equation}
\left\langle B\left| T_\nu ^\mu \right| B\right\rangle _{ren}=\left\langle
B\left| T_{\nu (gravitation)}^\mu \right| B\right\rangle +\left\langle
B\left| T_{\nu (boundary)}^\mu \right| B\right\rangle ,  \label{BB}
\end{equation}
\begin{equation}
\left\langle H\left| T_\nu ^\mu \right| H\right\rangle _{ren}=\left\langle
H\left| T_{\nu (gravitation)}^\mu \right| H\right\rangle +\left\langle
H\left| T_{\nu (boundary)}^\mu \right| H\right\rangle +\left\langle H\left|
T_{\nu (bath)}^\mu \right| H\right\rangle ,  \label{HH}
\end{equation}
\begin{equation}
\left\langle U\left| T_\nu ^\mu \right| U\right\rangle _{ren}=\left\langle
U\left| T_{\nu (gravitation)}^\mu \right| U\right\rangle +\left\langle
U\left| T_{\nu (boundary)}^\mu \right| U\right\rangle +\left\langle U\left|
T_{\nu (radiation)}^\mu \right| U\right\rangle ,  \label{UU}
\end{equation}
where $\left\langle T_{\nu (gravitation)}^\mu \right\rangle $ {and} $%
\left\langle T_{\nu (boundary)}^\mu \right\rangle $ denote the contribution
to vacuum zero point energy from the gravitation and boundary constraints,
respectively. $\left\langle T_{\nu (bath)}^\mu \right\rangle $ denotes the
contribution to vacuum zero point energy from the thermal bath at
temperature $T$. $\left\langle T_{\nu (radiation)}^\mu \right\rangle _{ren}$
denotes the contribution to vacuum zero point energy from Hawking radiation
at temperature $T$. Considering (\ref{BB}), (\ref{HH}) and (\ref{UU}), we
analyze the energy-momentum tensors for massless Dirac field in three vacuum
states (\ref{TB}), (\ref{v}) and (\ref{Z}) and find that 
\begin{eqnarray}
\left\langle B\left| T_{\nu (gravitation)}^\mu \right| B\right\rangle
&=&\left\langle H\left| T_{\nu (gravitation)}^\mu \right| H\right\rangle
=\left\langle U\left| T_{\nu (gravitation)}^\mu \right| U\right\rangle 
\nonumber \\
&=&\left( 
\begin{array}{cc}
\left\langle T_\alpha ^\alpha \right\rangle _{ren}-f^{-1}(r)H(r) & 0 \\ 
0 & f^{-1}(r)H(r)
\end{array}
\right) +f^{-1}\left( r\right) \frac{-1}{24\pi }  \nonumber \\
&&\times \left[ \frac{m^2}{\left( m+\sqrt{m^2-Q^2}\right) ^4}-\frac{2mQ^2}{%
\left( m+\sqrt{m^2-Q^2}\right) ^5}+\right.  \nonumber \\
&&\left. \frac{Q^4}{\left( m+\sqrt{m^2-Q^2}\right) ^6}\right] \times \left( 
\begin{array}{cc}
-1 & 0 \\ 
0 & 1
\end{array}
\right) ,  \label{Bbbbb}
\end{eqnarray}
\begin{eqnarray}
\left\langle B\left| T_{\nu (boundary)}^\mu \right| B\right\rangle
&=&\left\langle H\left| T_{\nu (boundary)}^\mu \right| H\right\rangle
=\left\langle U\left| T_{\nu (boundary)}^\mu \right| U\right\rangle 
\nonumber \\
&=&\frac \pi {24L^2}f^{-1}(r)\left( 
\begin{array}{cc}
-1 & 0 \\ 
0 & 1
\end{array}
\right) ,  \label{z111}
\end{eqnarray}
\begin{equation}
\left\langle H\left| T_{\nu (bath)}^\mu \right| H\right\rangle =f^{-1}(r)%
\frac{-1}{24\pi }\frac{\frac 12m^2-\frac 12Q^2}{\left( m+\sqrt{m^2-Q^2}%
\right) ^4}\left( 
\begin{array}{cc}
-1 & 0 \\ 
0 & 1
\end{array}
\right) ,  \label{z222}
\end{equation}
\begin{equation}
\left\langle U\left| T_{\nu (radiation)}^\mu \right| U\right\rangle
=f^{-1}(r)\left( 
\begin{array}{cc}
\frac 1{24\pi }\frac{\frac 14m^2-\frac 14Q^2}{\left( m+\sqrt{m^2-Q^2}\right)
^4} & -\frac 1{96\pi }\frac{m^2-Q^2}{\left( m+\sqrt{m^2-Q^2}\right) ^4} \\ 
\frac 1{96\pi }\frac{m^2-Q^2}{\left( m+\sqrt{m^2-Q^2}\right) ^4} & -\frac 1{%
24\pi }\frac{\frac 14m^2-\frac 14Q^2}{\left( m+\sqrt{m^2-Q^2}\right) ^4}
\end{array}
\right) .  \label{z333}
\end{equation}

Expression (\ref{Bbbbb}) shows that the contribution to vacuum zero point
energy from gravitation is the same in the three vacuum states. Expression (%
\ref{z111}) shows that the contribution to vacuum zero point energy from
boundary constraint is also the same in three vacuum states. Expressions (%
\ref{z222}) and (\ref{z333}) denote the contributions to vacuum zero point
energy from the thermal bath at temperature $T$ and Hawking radiation at
temperature $T$, respectively.

Now we can distinguish the different contributions from the gravitational
background (including the trace anomaly), boundaries, thermal bath and the
Hawking radiation. Obviously, the presence of boundary, i.e., the two
``parallel plates'', leads to a pressure 
\begin{equation}
P_b^{\left( 1,2\right) }=-\left\langle T_{r(boundary)}^r\right\rangle
=-f^{-1}(r_{1,2})\frac \pi {24L^2}.
\end{equation}
That is the Casimir force\ acting on two ``parallel plates'', appearing as
an attraction. It should be noted that the additional pressure created by
the other parts in Eqs. (\ref{BB})-(\ref{UU}) are the same from both sides
on the plates.

\section{Conclusion}

\label{Sec. 5}

There are many ways to group different cases of the Casimir effect: to give
some examples, zero temperature or nonzero temperature according to the
temperature of the field, static boundaries or moving boundaries according
to the motion state of boundaries, flat spacetime or curved spacetime
according to the background, scalar field, Dirac field, or gravitation
field, ...according to the spin of the field, and so on. To the best of our
knowledge, the issue of the Casimir effect for the massless Dirac field in
curved spacetime has not been studied before the present work.

From Refs. \cite{Davies,Setare,Chris,Vage}, we know that, in the
lower-dimensional case, if the energy-momentum tensor of a certain field
confined by an exterior boundary in the Minkowski spacetime can be obtained,
the energy-momentum tensor of the same field confined by the same boundary
in curved spacetime can be obtained too. Here the Casimir effect for a
massless Dirac field in the two-dimensional RN background has been studied.
The massless Dirac field is confined between two ``parallel plates''\
separated by a distance $L$, and there is no particle current through the
boundaries. We have derived the energy-momentum tensors, using only the
general properties of stress tensor, in the Boulware vacuum, the
Hartle-Hawking vacuum and the Unruh vacuum, respectively. We have found that
the Casimir energy-momentum tensors in the three vacuum states for a
massless Dirac field in the two-dimensional RN background when $Q\rightarrow
0$ are different from those for a massless scalar field in the
two-dimensional Schwarzschild background. However, we have found that under
the given constraints the Casimir force for a massless Dirac field in the
two-dimensional RN background is the same as that for a massless scalar
field in the two-dimensional Schwarzschild background \cite{Setare}.

It is interesting to compare some of the previous conclusions. In Ref. \cite
{Paola}, de Paola {\it et al.} studied the Casimir effect for a massless
Dirac field confined between two parallel plates in a $D$ dimensional flat
spacetime, and the massless Dirac field satisfies the Dirichlet boundary
condition. They obtained the Casimir energy in even dimensional spacetime 
\begin{equation}
\varepsilon _D\left( L\right) =-\frac{f\left( D\right) }{L^{D-1}}\Gamma
\left( \frac{1-D}2\right) \left( 2^{1-D}-1\right) \frac{B_D}D.  \label{eD}
\end{equation}
In $D=2$ dimension, the Casimir energy for the massless scalar field 
\begin{equation}
\varepsilon _2\left( L\right) =-\frac{f\left( 2\right) }{L^{2-1}}\Gamma
\left( -\frac 12\right) \left( 2^{-1}-1\right) \frac{B_2}2=-\frac \pi {24L}
\label{e2}
\end{equation}
in which $L$ is the distance between two `parallel plates'. In Ref. \cite
{Kimb}, Milton studied the Casimir effect for the massless scalar field
confined between two `parallel plates' in two dimensional flat spacetime,
and the confined massless scalar field satisfies the Dirichlet boundary
condition on the `parallel plates'. The author obtained the Casimir energy
for a massless scalar field 
\begin{equation}
E=-\frac \pi {24a},  \label{E123}
\end{equation}
in which $a$ is the distance between two `parallel plates'. Thus in the
two-dimensional case, the Casimir energy for the massless Dirac field
confined between two `parallel plates' is the same as the Casimir energy for
massless scalar field under the same condition. However, in four dimensional
flat spacetime, the Casimir energy for a massless Dirac field confined
between two `parallel plates' is \cite{Paola} 
\begin{equation}
\varepsilon _4\left( L\right) =-\frac{7\pi ^2}{2880L^3},  \label{e4}
\end{equation}
and the Casimir energy for a massless scalar field confined between two
`parallel plates' is \cite{Kimb} 
\begin{equation}
E_c\left( a\right) =-\frac{\pi ^2}{1440a^3}.  \label{Ecend}
\end{equation}

From the analysis above, it is only in higher-dimensional cases that the
massless Dirac field confined between two `parallel plates' produces a
different Casimir force on the plates from a scalar field in the same
condition. However, the Casimir effect for a massless Dirac field does
coincide with the Casimir effect for a massless scaler field in
two-dimensional spacetime, at least in the cases studied in the present work
and the cited references above.

\acknowledgments

The work was supported by the National Natural Science Foundation of China
(No. 10375008), and the National Basic Research Program of China
(2003CB716302). Gao wish to thank Chengzhou Liu for a useful discussion.

\end{document}